\documentclass[runningheads]{llncs}

\usepackage{amssymb}
\usepackage[table]{xcolor}
\usepackage{graphicx}
\usepackage{amsmath}
\usepackage{booktabs}
\usepackage{xcolor}
\usepackage{multirow}
\usepackage[hidelinks]{hyperref}

\usepackage[T1]{fontenc}
\usepackage{graphicx,verbatim}
\begin{document}
\title{INR meets Multi-Contrast MRI Reconstruction}

\author{
Natascha Niessen\inst{1,2}* \and
Carolin M. Pirkl\inst{2} \and
Ana Beatriz Solana\inst{2} \and
Hannah Eichhorn\inst{1,3} \and
Veronika Spieker\inst{1,3} \and
Wenqi Huang\inst{1} \and
Tim Sprenger\inst{2,4} \and
Marion I. Menzel\inst{2,5,6} \and
Julia A. Schnabel\inst{1,3,7} \\ on behalf of the PREDICTOM consortium
}

\authorrunning{Niessen et al.}

\institute{
School of Computation, Information and Technology, Technical University of Munich, Munich, Germany \and
GE HealthCare, Munich, Germany \and
Institute of Machine Learning in Biomedical Imaging, Helmholtz Munich, Neuherberg, Germany \and
Department of Clinical Neuroscience, Karolinska Institutet, Stockholm, Sweden \and
Technische Hochschule Ingolstadt, Ingolstadt, Germany \and
School of Natural Sciences, Technical University of Munich, Munich, Germany \and
King’s College London, London, United Kingdom\\
\email{*natascha.niessen@tum.de}
}

\maketitle              
\begin{abstract}
Multi-contrast MRI sequences allow for the acquisition of images with varying tissue contrast within a single scan. The resulting multi-contrast images can be used to extract quantitative information on tissue microstructure. To make such multi-contrast sequences feasible for clinical routine, the usually very long scan times need to be shortened e.g. through undersampling in k-space. However, this comes with challenges for the reconstruction. In general, advanced reconstruction techniques such as compressed sensing or deep learning-based approaches can enable the acquisition of high-quality images despite the acceleration.
In this work, we leverage redundant anatomical information of multi-contrast sequences to achieve even higher acceleration rates. We use undersampling patterns that capture the contrast information located at the k-space center, while performing complementary undersampling across contrasts for high frequencies. To reconstruct this highly sparse k-space data, we propose an implicit neural representation (INR) network that is ideal for using the complementary information acquired across contrasts as it jointly reconstructs all contrast images. We demonstrate the benefits of our proposed INR method by applying it to multi-contrast MRI using the MPnRAGE sequence, where it outperforms the state-of-the-art parallel imaging compressed sensing (PICS) reconstruction method, even at higher acceleration factors.
Our code is available at \href{https://github.com/compai-lab/2025-miccai-niessen}{https://github.com/compai-lab/2025-miccai-niessen}.


\end{abstract}

\section{Introduction}

Quantitative assessment of brain tissue microstructure is essential for identifying novel biomarkers e.g. associated with neurodegenerative diseases such as multiple sclerosis and Alzheimer’s disease. Multi-contrast magnetic resonance imaging (MRI) sequences~\cite{qMRI} provide a rich set of comprehensive tissue contrasts and enable extraction of quantitative information within a single scan by acquiring images with varying contrast, through changes in experimental parameters such as inversion time or echo time. However, these sequences are often time-consuming, limiting clinical feasibility and increasing the risk of motion artifacts. To achieve shorter scan times, one approach is to acquire fewer k-space samples. However, undersampling k-space data is known to produce artifacts and low SNR, oftentimes rendering the MR images clinically meaningless.

Many multi-contrast MRI sequences have in common, that the anatomy of the object being scanned remains constant, disregarding potential motion, field inhomogeneity etc., while the contrast changes with the experimental parameters. In this work, we leverage this property to achieve high acceleration rates for multi-contrast MRI sequences in neuro applications through complementary undersampling in k-space along the additional dimension.

Deep learning-based reconstruction show promising results for accelerated MRI~\cite{Hammernik2023,Heckel2024}. However, supervised approaches require large datasets for training and lack generalizability e.g. with respect to varying MRI sequences. In contrast, self-supervised approaches reconstruct the MR image only based on the acquired data itself which is ideal for multi-contrast settings where training data is very scarce. One prominent self-supervised architecture that emerged from the field of computer vision is implicit neural representation (INR)~\cite{nerf} that allows to represent images or objects continuously with the help of a small neural network and an input encoding. The INR architecture enables a patient-specific, compact representation of high-dimensional multi-contrast images with fewer network parameters, capturing shared information and evolving contrast in a compressed form. Our approach is inspired by recent work leveraging INRs for dynamic MRI~\cite{SubspaceINR,Spatiotemporal}. In applications such as cardiac cine MRI, the heart as an object remains the same, while undergoing temporal deformation due to cardiac motion. Analogously, in the neuroimaging context we target, multi-contrast sequences yield a series of images in which the brain’s anatomy remains the same, while the image contrast evolves across acquisitions.

In this work, we propose an INR based reconstruction framework for accelerated multi-contrast MRI acquisitions. We evaluate our method through retrospective variable density Poisson disk undersampling of fully sampled k-space data of the brain, acquired with the MPnRAGE sequence~\cite{MPnRAGE}, that provides multiple inversion contrast images. Beyond the rich contrast information provided by MPnRAGE, the resulting images can be used for fitting a quantitative T1 map or to investigate tissue nulling, similar to fluid suppression in FLAIR~\cite{FLAIR}.

\section{Related work}

\subsubsection{INRs for MRI reconstruction}

Recent works use INRs for undersampled MRI reconstruction either directly in k-space~\cite{NIK,ICoNIK} or, like in our work, in image domain~\cite{SubspaceINR,Spatiotemporal,TabitaINR,FourierINR,IMJENSE,MP-qMRI_INR,NeRP}. However, most of these methods focus on dynamic MRI ~\cite{FourierINR} or more specifically cardiac cine MRI with radial acquisitions~\cite{SubspaceINR,Spatiotemporal,TabitaINR}. Lao et al. target multi-parametric quantitative MRI~\cite{MP-qMRI_INR} and for Shen et al.~\cite{NeRP} a prior reconstructed MR image is required.

\subsubsection{Deep learning-based multi-contrast MRI reconstruction}
Joint multi-contrast reconstruction with complementary undersampling is explored for a combination of T1w, T2w and T2-FLAIR images with an unrolled neural network by Polak et al.~\cite{JointVar}. A method proposed by Lei et al.~\cite{DecompVarNet} guides the unrolled reconstruction of T2w images with the help of fully sampled proton density (PD) reference images. Both approaches are supervised methods requiring large training datasets. Moreover, these methods may be susceptible to image registration effects, a limitation that our approach circumvents by acquiring data simultaneously.

\section{Method}

\subsection{Multi-contrast MRI reconstruction}

In multi-contrast MRI reconstruction, the goal is to reconstruct multiple images $\mathbf{d} \in \mathbb{C}^{ (V_y \times V_z) \times N}$ of the same object, acquired with a different contrast $n = 1,...,N$. In this work, we consider 2D slices of the object with image dimensions $V_y$ and $V_z$ as $k_x$ corresponds to the readout direction. In MRI, noisy linear measurements $\mathbf{D}_c \in \mathbb{C}^{ (V_y \times V_z) \times N}$ are acquired according to the forward model

\begin{equation}
\mathbf{D}_c = \mathbf{M} \mathbf{F} \mathbf{S}_c \mathbf{d} + \mathbf{e}_c
\end{equation}

\noindent at the receiver coils $c=1,..,C$, where $\mathbf{M}$ is a binary undersampling mask, $\mathbf{F}$ is the 2D Fourier matrix and $\mathbf{S}_c$ is a diagonal matrix containing the coil sensitivity maps of the receiver coil $c$. The measurements are distorted by additive Gaussian noise $\mathbf{e}_c$.

\subsubsection{Complementary undersampling}

\begin{figure}[h]
\includegraphics[width=\textwidth]{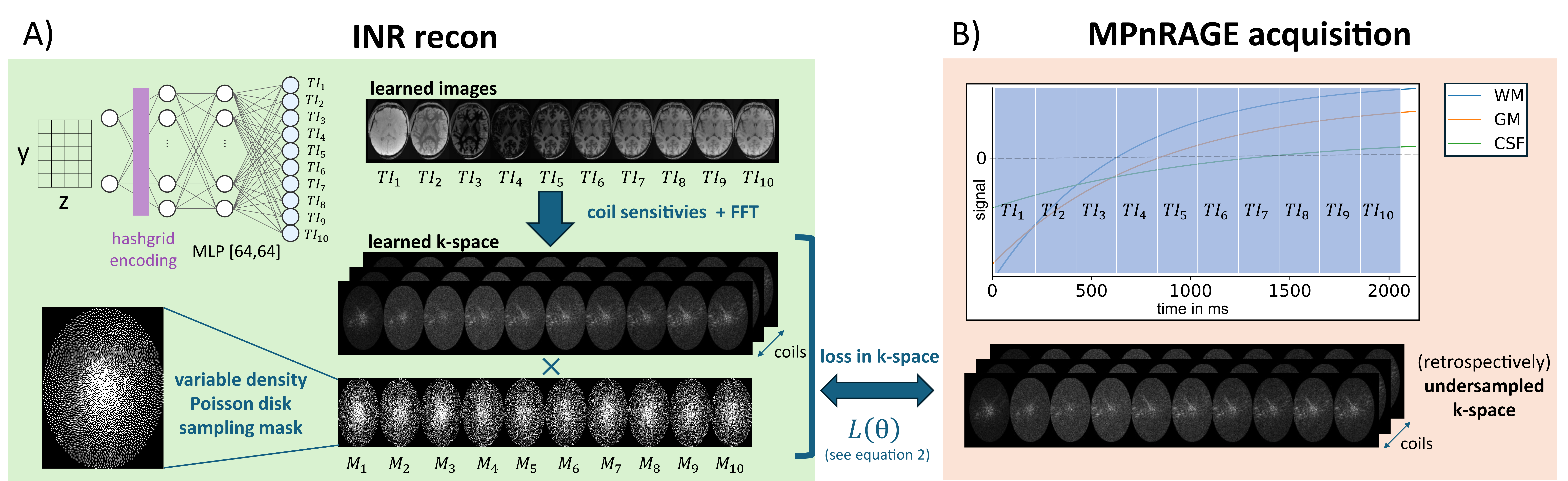}
\caption{A) Our proposed INR reconstruction framework for accelerating multi-contrast MRI sequences through complementary undersampling in k-space demonstrated for the MPnRAGE sequence. B) Exemplary inversion recovery signal curves for white matter (WM), gray matter (GM) and cerebrospinal fluid (CSF). The zero-crossings of the different tissues are visible in the TI contrast images in A (learned images).} \label{fig1}
\end{figure}

In our multi-contrast setup, we consider the undersampling mask $\mathbf{M} = [\mathbf{M}_1,...,\mathbf{M}_N]$ to be different for each of the $n = 1,...,N$ acquired contrasts. To sufficiently encode the contrast evolution, we densely sample the k-space center, while the outer k-space is sampled sparsely and complementarily across contrasts. More specifically, we use variable density Poisson disk sampling, a strategy that places sampling points in k-space in a pseudo-random yet spatially balanced manner (see Figure~\ref{fig1}A). This introduces incoherent, noise-like aliasing artifacts, which are effectively mitigated by deep learning-based reconstruction methods~\cite{Heckel2024}. By using different random seeds for each contrast, we generate complementary sampling patterns for each multi-contrast data which collectively enhance k-space coverage.

\subsection{Implicit Neural Representation}

For reconstruction, we represent the images with an INR network, \mbox{$\mathbf{d} = G_\theta(\mathbf{y},\mathbf{z})$} with model parameters $\theta$, similarly to Huang et al.~\cite{SubspaceINR}, but jointly embed all multi-contrast images. The INR network learns a continuous mapping \mbox{$G_\theta(\cdot) : \mathbb{R}^2 \rightarrow \mathbb{C}^N$} from the image coordinates \mbox{$(\mathbf{y},\mathbf{z}) \in \mathbb{R}^2$} to the complex voxel-wise signal evolution with $N$ different contrasts. By jointly reconstructing all contrast images, the INR enforces data consistency for each individual contrast. It leverages the shared anatomical structure present across contrasts, made accessible through the complementary undersampling strategy. Although the inference operates on a voxel-wise basis, the INR captures the entire encoded spatio-temporal information since all contrast images are reconstructed jointly.

\subsubsection{Loss function in k-space with distance weighting}

We constructed the loss function such that it promotes each individual contrast image to be in accordance with its acquired k-space data. To calculate the loss at each epoch, we multiply the learned images with the coil sensitivity maps and apply a fast fourier transform (FFT) to the resulting coil images. Hereby, we obtain what we refer to as the “learned k-space” (see Figure~\ref{fig1}A). For data consistency, we apply the complementary undersampling masks, to the learned k-space data. Finally, we compare the resulting masked learned k-space data to the acquired k-space data by calculating the mean-squared error 

\begin{equation}
L(\theta) = \sum_{c=1}^{C} \Big\| W(\mathbf{k}_y,\mathbf{k}_z) \cdot \left[ \mathbf{M} \mathbf{F} \mathbf{S}_c G_\theta(\mathbf{y},\mathbf{z}) - \mathbf{D}_c \right] \Big\|_2^2
\end{equation}

\noindent using weights proportional to the Euclidean distance from the k-space center

\begin{equation}
W(\mathbf{k}_y, \mathbf{k}_z) = \sqrt{\mathbf{k}_y^2 + \mathbf{k}_z^2} + 1
\end{equation}

\noindent These weights ensure that all k-space samples contribute to the loss function, despite the high-dynamic range between high intensities at the k-space center and low intensities at higher frequencies.

\subsubsection{INR architecture}
The INR network that we use for reconstruction consists of a multi-layer perceptron (MLP) with two hidden linear layers of 64 neurons each, using ReLu as activation function. The relatively low dimensionality of the MLP is possible thanks to the hashgrid encoding~\cite{hashgrid} of the input coordinates that has proven to contribute to the reconstruction performance of the INR~\cite{SubspaceINR,Spatiotemporal}. We use the same hyperparameters for all datasets, to show the generalizability of our proposed method that does not require any subject-specific hyperparameter tuning. The INR is optimized per slice, where the final epoch of training simultaneously yields the inference result.

\subsection{Data acquisition using the MPnRAGE sequence}

While MPRAGE is the state-of-the-art sequence to obtain one single T1-weighted image, MP\textbf{n}RAGE allows to obtain \textbf{N} inversion time (TI) images that correspond to different T1 weightings and hence varying tissue contrasts. The scan duration of state of the art MPnRAGE sequences of around 15 min is considered clinically infeasible, but a recent work investigates the sparsity of the sequence through subspace compression and thus proves that there is a potential for acceleration in the TI dimension~\cite{MPnRAGE_ISMRM2025}.

The TI images are acquired along the inversion recovery curve, following an inversion preparation module. Figure~\ref{fig1}B visualizes the acquisition for $N = 10$, showing that k-space data acquired within a certain TI interval is binned, to form one TI image. The inversion recovery is repeated several times in order to acquire the entire 3D+TI k-space in an iterative way. This segmented k-space readout is performed in 3D with $k_x$ as readout direction. This enables us to apply a one dimensional FFT in $k_x$ direction and reconstruct each axial slice individually. To simulate the acceleration, we retrospectively undersample in the $k_y-k_z$ plane corresponding to the axial view. The fully sampled Cartesian 3D MPnRAGE sequence was implemented in KSFoundation~\cite{ksfoundation}. Healthy volunteer data from three subjects were acquired on a DISCOVERY™ MR750w (GE HealthCare, Waukesha, USA) in the sagittal scan orientation with the following parameters: FOV 24$\times$24$\times$18 cm², matrix size 160$\times$160$\times$120, for 10 TIs (\mbox{$TI_1 = 26\, \text{ms}$}, \mbox{$\Delta_{TI} = 249.05\,\text{ms}$}), imaging TR 4.98 ms, imaging flip angle 4°, MPnRAGE TR 2.7 s, acquisition time 13.47 min.

\subsubsection{Analysis}

For evaluation, we apply a brain mask calculated using the FSL toolbox~\cite{FSL} and perform percentile normalization jointly across all contrasts and slices~\cite{motionIQ}. The resulting masked and normalized data is then used to compute the structural similarity index measure (SSIM) and the peak signal-to-noise ratio (PSNR) relative to the fully sampled reference.

\begin{figure}[h]
\includegraphics[width=\textwidth]{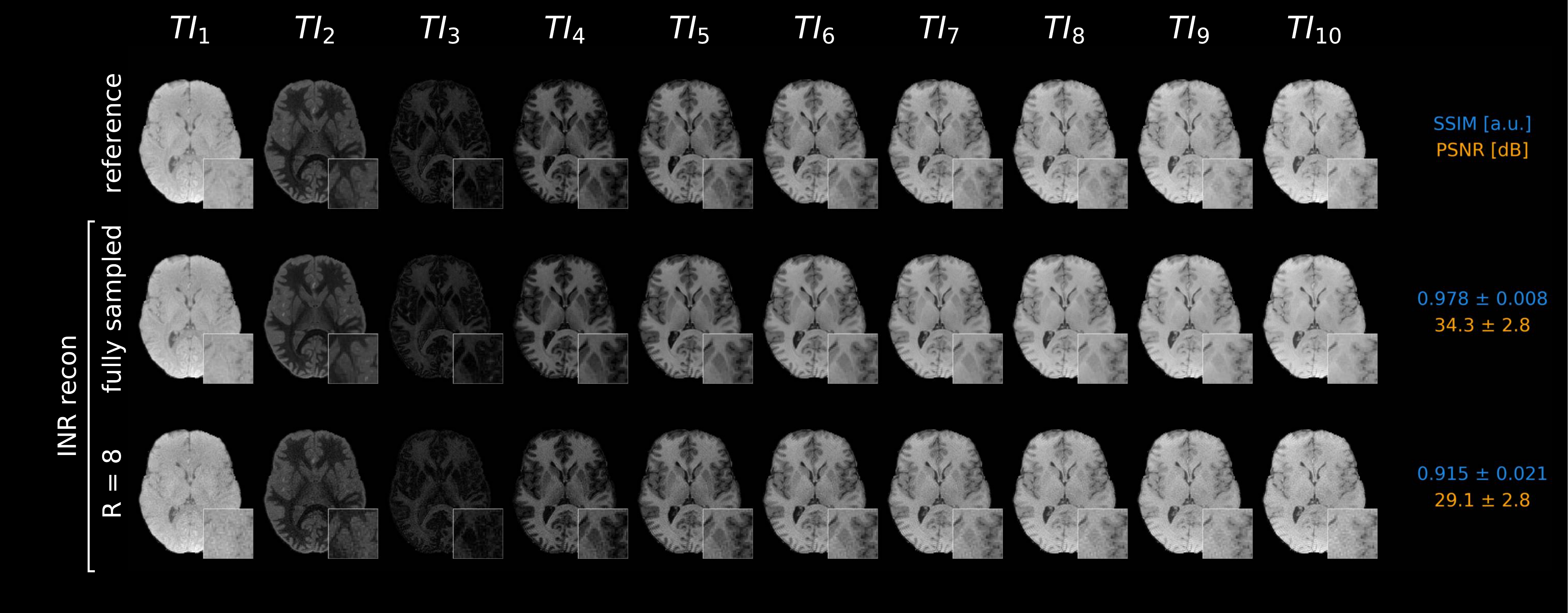}
\caption{$N = 10$ inversion time contrast images acquired with the MPnRAGE sequence, reconstructed for fully sampled data with an inverse FFT (reference) and using our joint INR reconstruction for fully sampled and undersampled data (R = 8). The metrics correspond to the shown slice and were averaged over all contrasts.} \label{fig2}
\end{figure}

\section{Experiments and results}

\subsubsection{Denoising effect: Fully-sampled INR recon vs. reference image}
First, we evaluate the INR framework for fully sampled data. To obtain a reference image, we reconstruct the same fully sampled data with a standard inverse FFT using coil sensitivity maps.
In Figure~\ref{fig2} we compare the reconstructions of the 10 multi-contrast images with the proposed INR framework, based on the fully sampled k-space data, to the reference images. The INR-obtained images preserve fine anatomical details, as evidenced by a high SSIM of 0.978 $\pm$ 0.008  and a PSNR of 34.3 $\pm$ 2.8 dB, averaged over all slices and contrast images. The denoising effect due to the inherent regularization of the INR becomes clearly visible in the zoomed insets (Figure~\ref{fig2}). The unified training-inference process takes approximately 0.8 seconds per slice.

\subsubsection{Acceleration through undersampling: INR recon vs. PICS}

\begin{figure}[h] \includegraphics[width=\textwidth]{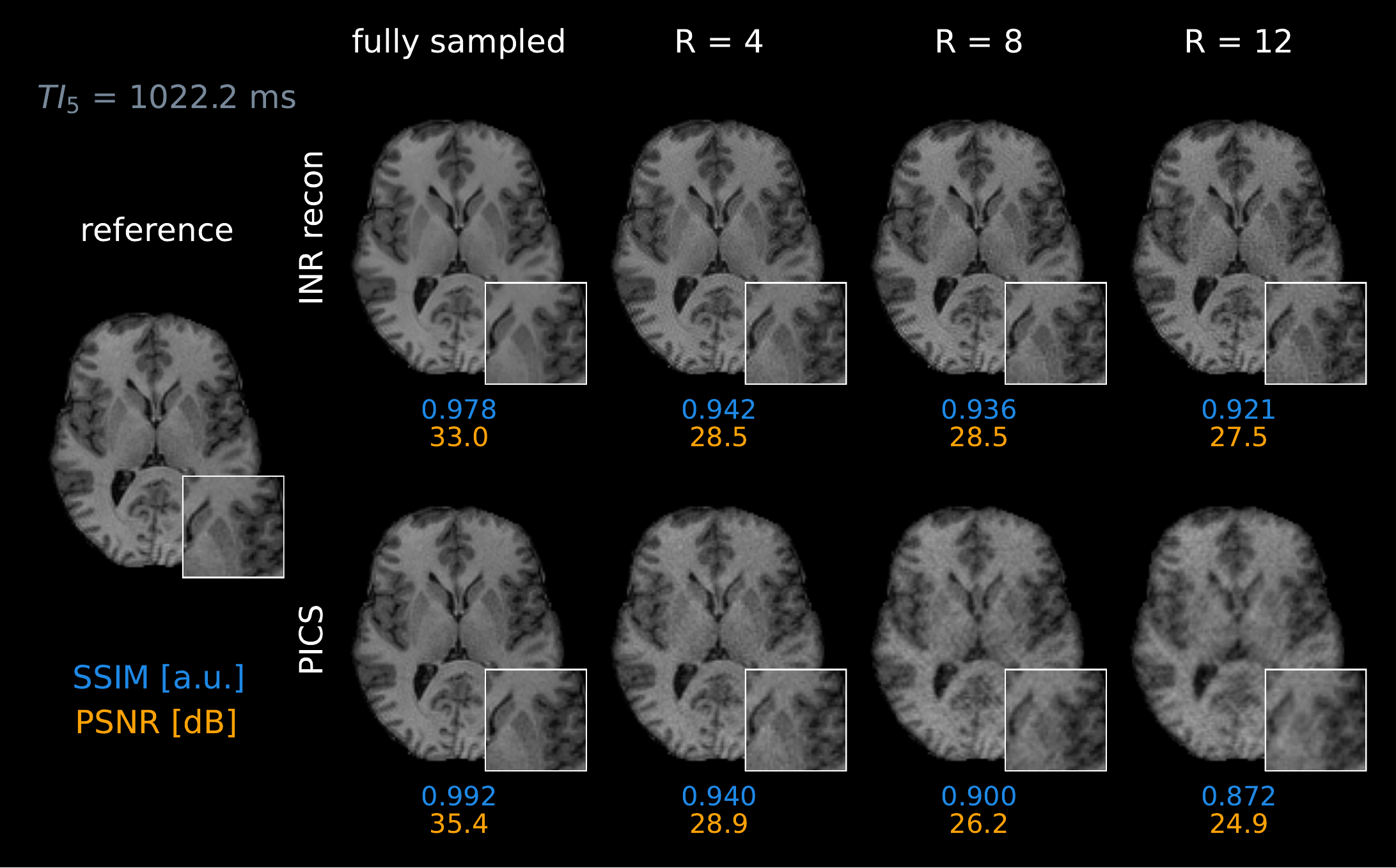} \caption{Exemplary reconstructed contrast images ($TI_5$ = 1022.2 ms) resulting from our joint INR reconstruction and PICS, for fully sampled data and various acceleration factors~R. An inverse FFT of the fully sampled data serves as reference image. The metrics correspond to the shown slice and contrast.} \label{fig3} \end{figure}

Next, we retrospectively undersample the fully sampled k-space data to simulate the acceleration of the multi-contrast sequence at R = 4, 8 and 12. We compare our method to PICS~\cite{PICS}, a classical state-of-the-art approach for mitigating incoherent artifacts, such as those introduced by our random undersampling pattern.
The SSIM and PSNR metrics in table~\ref{tab1} demonstrate that our proposed INR reconstruction framework performs comparably to PICS in the fully sampled case. For accelerations up to R = 12, the SSIM and PSNR of the INR method decline only slightly, indicating strong robustness to undersampling. In contrast, PICS shows a much stronger drop in both metrics, reflecting a significant degradation in image quality which can be seen in the example images in Figure~\ref{fig3}. The proposed INR framework preserves fine structural details even at high acceleration rates.

\subsubsection{Cross-Plane Contrast Continuity and Spatial Alignment}
\begin{figure}
\includegraphics[width=\textwidth]{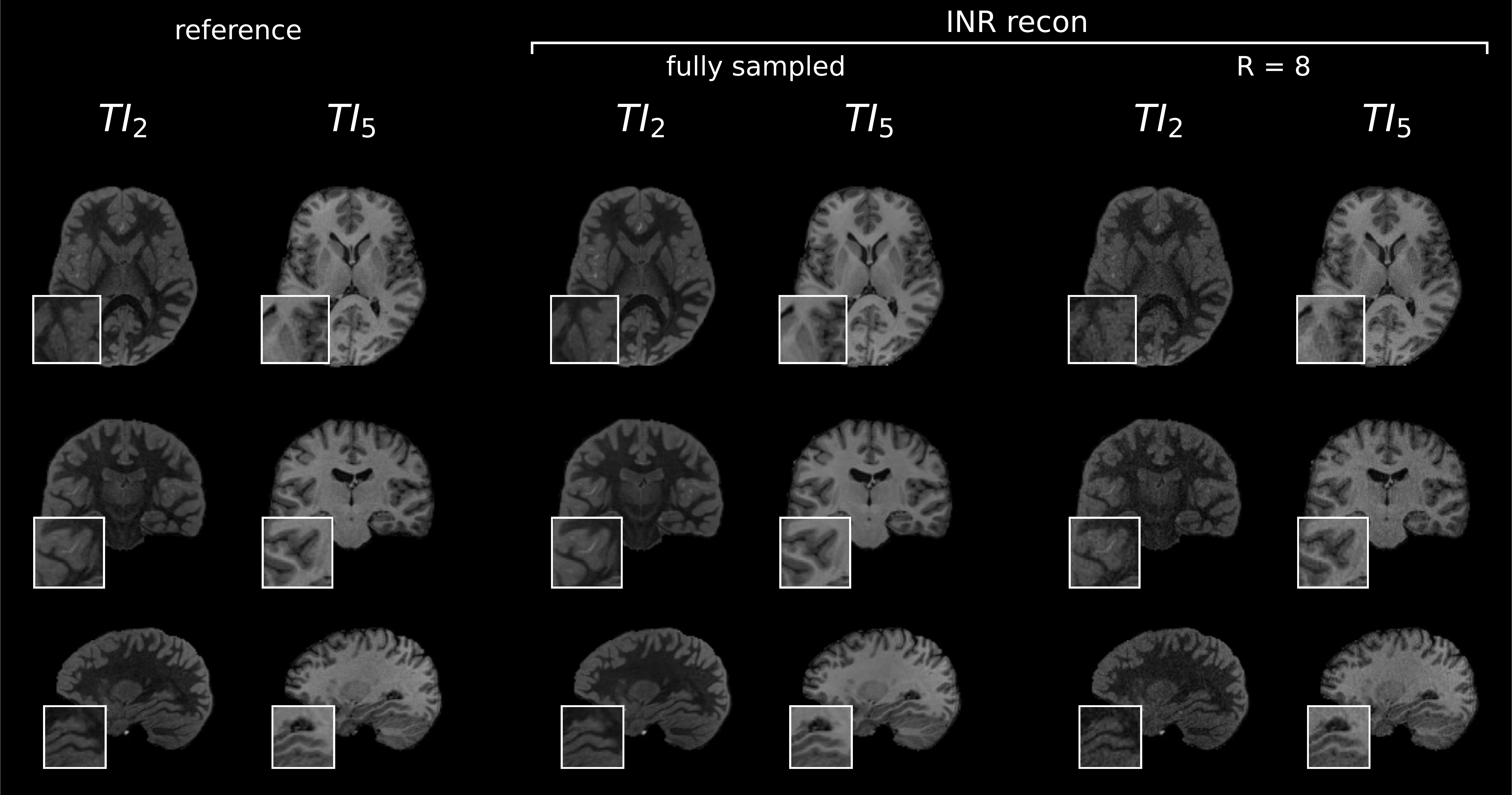}
\caption{Reconstructed axial slices, as well as coronal and sagittal views extracted from the stack of axial slices obtained using our joint INR reconstruction with fully sampled and undersampled (R = 8) data, for two exemplary inversion time (TI) contrasts. An inverse FFT of the fully sampled data serves as reference image.} \label{fig4}
\end{figure}

As previously mentioned, the reconstruction is performed independently for each axial slice. In Figure~\ref{fig4} we visualize sagittal and coronal planes extracted from the reconstructed axial slice stack to show the cross-plane contrast continuity achieved with our proposed INR framework. Additionally, the individually reconstructed slices are accurately aligned in space, which demonstrates that our approach maintains the global coordinate space through the data consistency enforced by the loss function in k-space.

\begin{table}[h]
\caption{SSIM [a.u.] and PSNR [dB] across different acceleration rates for INR reconstruction and PICS, averaged over all volunteers, contrasts, and slices.}
\label{tab1}
\centering
\begin{tabular}{|c|c|c|c|c|c|}
\hline
\textbf{} & \textbf{} & \textbf{Fully Sampled} & \textbf{R = 4} & \textbf{R = 8} & \textbf{R = 12} \\
\hline
\multirow{2}{*}{\textbf{INR recon}} & SSIM & 0.981 $\pm$ 0.002 & \textbf{0.959 $\pm$ 0.005} & \textbf{0.947 $\pm$ 0.006} & \textbf{0.935 $\pm$ 0.008} \\
 & PSNR & 33.9 $\pm$ 1.4 & 31.5 $\pm$ 0.7 & \textbf{30.6 $\pm$ 0.9} & \textbf{30.0 $\pm$ 0.6} \\
\hline
\multirow{2}{*}{\textbf{PICS}} & SSIM & \textbf{0.987 $\pm$ 0.001} & 0.953 $\pm$ 0.004 & 0.924 $\pm$ 0.005 & 0.903 $\pm$ 0.006 \\
 & PSNR & \textbf{36.0 $\pm$ 0.9} & \textbf{31.8 $\pm$ 0.3} & 29.4 $\pm$ 0.1 & 28.0 $\pm$ 0.2 \\
\hline
\end{tabular}
\end{table}

\section{Discussion and conclusion}

In this work, we propose a joint INR-based reconstruction framework that leverages shared information across multi-contrast images to produce high-quality reconstructions at high acceleration factors. The proposed method was evaluated using an inversion recovery contrast sequence and based on the shared underlying MR physics, is anticipated to generalize effectively to other multi-contrast sequences, such as T2-weighted imaging. In the fully sampled case, our method performs comparably to PICS. The slight performance gap is likely due to the INR’s inherent regularization, which promotes spatial smoothness and therefore acts as a denoiser. This denoising has a negative effect on the metrics, as the reference images themselves contain noise. The reference image for one of the volunteers showed blurring, potentially introduced by slight motion. Our INR framework was able to mitigate parts of this blurring, as observed in the fully sampled case and even at an acceleration factor of R = 4 (Supplementary Material, Figure 5). This demonstrates the potential of our joint INR-based reconstruction, which leverages data acquired at different time points across all contrasts by embedding them within a shared network.

At high acceleration factors up to R = 12, which corresponds to a shortening of the scan time from initially 13.47 min to 1.12 min, our joint INR framework significantly outperforms PICS both visually and quantitatively. The presented metrics are very promising regarding the robustness of our joint INR framework to increasingly undersampled data, from which we conclude that even higher accelerations could be achieved. However, for Poisson disk sampling, generating even more sparse sampling patterns becomes algorithmically complex, which is why for future work we also consider exploring other undersampling strategies,  potentially even in 3D, or optimizing the complementary sampling instead of relying on varying random seeds. The INR reconstructions tend to exhibit increased Gibbs ringing at an acceleration factor of R = 4. A deeper investigation into how the choice of undersampling trajectory interacts with INR-based reconstruction methods to reduce such artifacts is planned as part of our ongoing research efforts. In future work, we aim to implement the most promising sampling strategy in a prospectively accelerated acquisition and compare the reconstruction also to other deep learning-based reconstructions. This will enable high-quality imaging and extremely efficient acquisition of multi-contrast datasets for improved diagnostics and biomarker discovery.

\textbf{Conclusion} We introduce a joint INR-based reconstruction framework with complementary undersampling for accelerated multi-contrast MRI acquisitions. Our method was validated through retrospective undersampling using variable-density Poisson disk patterns applied to fully sampled in vivo data acquired with the MPnRAGE sequence, providing multiple inversion contrast images. The resulting images demonstrate that our approach enables high-quality reconstructions at substantially reduced scan time.

\subsubsection{Acknowledgments}
This work is supported by the DAAD programme Konrad Zuse Schools of Excellence in Artificial Intelligence and the Munich Center for Machine Learning , both sponsored by the Federal Ministry of Research, Technology and Space. PREDICTOM is supported by the Innovative Health Initiative Joint Undertaking (IHI JU), under Grant Agreement No 101132356. JU receives support from the European Union’s Horizon Europe research and innovation programme, COCIR, EFPIA, EuropaBio, MedTechEurope and Vaccines Europe. The UK participants are supported by UKRI Grant No 10083467 (National Institute for Health and Care Excellence), Grant No 10083181 (King's College London), and Grant No 10091560 (University of Exeter). University of Geneva is supported by the Swiss State Secretariat for Education, Research and Innovation Ref No 113152304. See \href{https://www.ihi.europa.eu}{www.ihi.europa.eu} for more details.
\newline
\textbf{Disclosure of Interests} NN, CMP, ABS, TS, and MIM are employees of GE HealthCare, Munich. All other authors declare that they do not have any financial or non-financial conflict of interests.

%
%
%
%

\newpage

\section*{Supplementary material}

\begin{figure}[h]
\includegraphics[width=\textwidth]{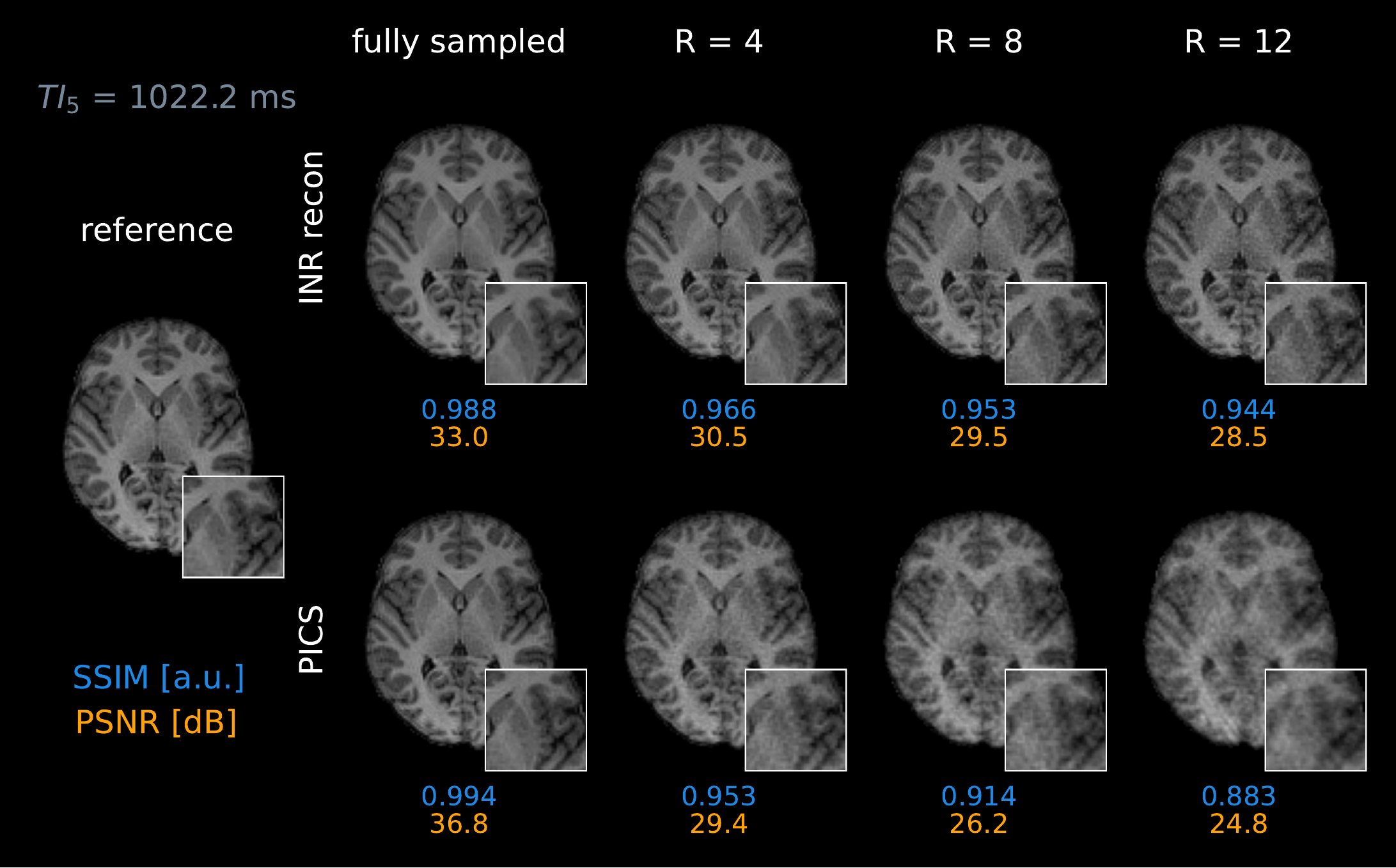}
\caption{Same visualization as Figure 3 but for another volunteer that showed more motion in the form of blurring in the reference image. For this volunteer, the contrast images resulting from our proposed INR framework indicate that our joint INR reconstruction has the potential to remove motion artifacts.} \label{fig5}
\end{figure}

\end{document}